%A SET OF MACROS FOR WRITING PAPERS.
%
%\today GIVES TODAY'S DATE.
%
\def\today{\ifcase\month\or January\or February\or March\or April\or May\or
June\or July\or August\or September\or October\or November\or December\fi
\space\number\day, \number\year}
%
%\note{footnote} GIVES SEQUENTIALLY NUMBERED FOOTNOTES.
%
\newcount\notenumber

\def\note{\global\advance\notenumber by 1 \footnote{$^{\the\notenumber}$}}
%
%\numbereq SEQUENTIALLY NUMBERS EQUATIONS ON THE RIGHT (number)
%
\newif\ifsectionnumbering
\newcount\eqnumber
\def\cleareqnumber{\eqnumber=0}
\def\numbereq{\global\advance\eqnumber by 1
\ifsectionnumbering\eqno(\the\secnumber.\the\eqnumber)
\else\eqno(\the\eqnumber) \fi}
\def\eqalinno{{\global\advance\eqnumber by 1}
\ifsectionnumbering(\the\secnumber.\the\eqnumber)\else(\the\eqnumber)\fi}
\def\name#1{\ifsectionnumbering\xdef#1{\the\secnumber.\the\eqnumber}
\else\xdef#1{\the\eqnumber}\fi}

\sectionnumberingtrue
%
%\ref{\name} GIVES SEQUENTIALLY NUMBERED REFERENCES [number], AND ASSIGNS
%THAT NUMBER TO A MACRO \name AND WRITES REF. TO FILE 1.
%
\newcount\refnumber

\immediate\openout1=refs.tex
\immediate\write1{\noexpand\frenchspacing}
\immediate\write1{\parskip=0pt}
\def\ref#1#2{\global\advance\refnumber by 1%
[\the\refnumber]\xdef#1{\the\refnumber}%
\immediate\write1{\noexpand\item{[#1]}#2}}
\def\tie{\noexpand~}

%
% NEW SECTION: \newsection The Method. (terminate with a .)
%
\font\twelvebf=cmbx10 scaled \magstep1
\newcount\secnumber

\def\newsection#1.{\ifsectionnumbering\cleareqnumber\else\fi%
	\global\advance\secnumber by 1%
	\bigbreak\bigskip\par%
	\line{\twelvebf \the\secnumber. #1.\hfil}\nobreak\medskip\par}
%
%
%\Box GIVES WAVE OPERATOR, OR LAPLACIAN
%
\def \sqr#1#2{{\vcenter{\vbox{\hrule height.#2pt
	\hbox{\vrule width.#2pt height#1pt \kern#1pt
		\vrule width.#2pt}
		\hrule height.#2pt}}}}

%
%
%\twocolumns GIVES TWO-COLUMN OUTPUT
%
\newdimen\fullhsize
\def\fiddle{\fullhsize=6.5truein \hsize=3.2truein}
\def\fullline{\hbox to\fullhsize}
\def\mkhdline{\vbox to 0pt{\vskip-22.5pt
	\fullline{\vbox to8.5pt{}\the\headline}\vss}\nointerlineskip}
\def\mkftline{\baselineskip=24pt\fullline{\the\footline}}
\let\lr=L \newbox\leftcolumn
\def\twocolumns{\fiddle
	\output={\if L\lr \global\setbox\leftcolumn=\columnbox
		\global\let\lr=R \else \doubleformat \global\let\lr=L\fi
		\ifnum\outputpenalty>-20000 \else\dosupereject\fi}}
\def\doubleformat{\shipout\vbox{\mkhdline
		\fullline{\box\leftcolumn\hfil\columnbox}
		\mkftline} \advancepageno}
\def\columnbox{\leftline{\pagebody}}
\magnification=1200
\def\pr#1 {Phys. Rev. {\bf D#1\tie }}
\def\pre#1 {Phys. Rep. {\bf #1\tie}}
\def\pe#1 {Phys. Rev. {\bf #1\tie}}
\def\pl#1 {Phys. Lett. {\bf #1B\tie }}
\def\prl#1 {Phys. Rev. Lett. {\bf #1\tie }}
\def\np#1 {Nucl. Phys. {\bf B#1\tie }}
\def\ap#1 {Ann. Phys. (NY) {\bf #1\tie }}
\def\cmp#1 {Commun. Math. Phys. {\bf #1\tie }}
\def\imp#1 {Int. Jour. Mod. Phys. {\bf A#1\tie }}
\def\mpl#1 {Mod. Phys. Lett. {\bf A#1\tie}}
%% poor man's black board bold
\def\BbbZ{{}\kern+1.6pt\hbox{$I$}\kern-7.5pt\hbox{$Z$}}

\def\tie{\noexpand~}

\parskip=15pt plus 4pt minus 3pt
\headline{\ifnum \pageno>1\it\hfil  Symmetries of 2-d Gravity
	$\ldots$\else \hfil\fi}
\font\title=cmbx10 scaled\magstep1
\font\tit=cmti10 scaled\magstep1
\footline{\ifnum \pageno>1 \hfil \folio \hfil \else
\hfil\fi}
\raggedbottom
\rightline{\vbox{\hbox{RU96-15-B}}}
\vfill
\centerline{\title SYMMETRIES OF 2-D
GRAVITY}
\vfill
{\centerline{\title Ioannis Giannakis$^{(a)}$ \footnote{$^{\dag}$}
{\rm e-mail: giannak@theory.rockefeller.edu}}
}
\medskip
\centerline{$^{(a)}${\tit Physics Department, Rockefeller
University}}
\centerline{\tit 1230 York Avenue, New York, NY
10021-6399}
\vfill
\centerline{\title Abstract}
\bigskip
{\narrower\narrower
Two-dimensional gravity in the light-cone gauge was shown
to exhibit an underlying $sl(2, R)$ current algebra. It is
the purpose of this note to offer a possible explanation
about the origin of this important algebra. The essential
point is that two-dimensional gravity is governed by a
topological field theory. The gauge group is $sl(2, R)$
and it is this enhanced gauge group that yields Polyakov's
current algebra.\par}
\vfill\vfill\break

\newsection Introduction.

Gravity in two-dimensional space-times has received extensive
scrutiny in recent years, both because of its importance in string
theory, where it plays a crucial r\^ole in world-sheet dynamics,
and as a tractable toy model for a quantum theory of gravity in
higher dimensions. As is usually the case for an effective toy
model, the dynamics of two-dimensional gravity is just non-trivial
enough to be interesting, while remaining largely tractable.

At first sight, two-dimensional general relativity would appear to
be a {\it too\/} trivial topological theory, since the Einstein-Hilbert
Lagrangean density is a total divergence, and the space-time metric
is a gauge artifact devoid of classical dynamics. Quantum mechanically,
of course, this is not true; except when coupled to conformal matter
with an appropriate central charge, quantum effects induce non-trivial
dynamics for the metric modulo coordinate transformations. In this
way, we are led  naturally to versions of two-dimensional scalar-tensor
gravity in which non-trivial equations of motion specify the scalar
curvature to be a (cosmological) constant. In conformal gauge this
leads to Liouville theory, but much greater progress has come from
Polyakov's introduction of lightcone gauge, in which the only degree
of freedom is the $ g_{++}$ component of the metric.  This is a
natural gauge choice because $ \det{g} = 1$, and all geometrical
quantities (Christoffel symbols, curvatures) are polynomial in  $
g_{++}$ and its derivatives. However, the real power of this gauge
choice lies in Polyakov's discovery of an $ sl(2,R)$ current algebra
within the algebra of observables \ref{\pol}{
A. M. Polyakov \mpl2 (1987), 893.}. It is this current algebra that
makes the theory tractable, but its appearance would appear to be
entirely mysterious---unmotivated by any principle that goes into
the formulation of the theory. Several proposals towards unraveling
its origin have been made
in the literature \ref{\duo}{M. Bershadsky and H. Ooguri, \cmp126
(1989), 49; A. Alekseev and S. Shatashvili, \np323
(1989), 719; A. Chamseddine and M. Reuter, \np317 (1989), 757;
E. Argyres, C. Papadopoulos and E. Floratos \pl234
(1990), 304; J. Helayel-Neto, S. Mokhtari and A. Smith, \pl236 (1990),
12; E. Egorian and R. Manvelian, \mpl5 (1990), 2371; E. Abdalla,
M. Abdalla, J. Gamboa and A. Zadra, \pl273 (1991), 222;
A. Smallagic and E. Spallucci, \pl317 (1993), 526;
P. Forgacs, A. Wipf, J. Balog, L. Feher and L. O' Raifeartaigh,
\pl227 (1989), 214; P. Schaller and T. Strobl, \pl337 (1994), 266;
N. Mohammedi, \mpl5 (1990), 1251.}.
It is our purpose in this note to offer an alternative
explanation about the origin of this important algebra.

The essential point is that, despite appearances, this theory too
is governed by a topological field theory, this time of Chern-Simons
type ({\it  i.e.} the solutions are flat connections). The non-trivial
dynamics emerges through a projection, rather in the same way that
solutions of the self-dual Yang-Mills equations emerge from a
projection in twistor or harmonic space. The gauge group of this
Chern-Simons theory contains the local Lorentz group, but is larger;
in two dimensions it is just $ sl(2,R)$, and we shall show that it
is this enhanced gauge group that yields Polyakov's current algebra.
This gauge theoretic formulation of two-dimensional gravity has
received some attention in the literature, but is in fact a
two-dimensional version of a four-dimensional construction first
described in a little noticed paper of Pagels \ref{\page}{
H. Pagels, \pr29 (1984), 1690.}, in which
some of the basic ideas of
topological field theory \ref{\witee}{E. Witten,
\cmp117 (1988), 353; \np311 (1988), 46.} were
introduced. Related ideas
can also be found in \ref{\cham}{A. H. Chamseddine and
P. C. West, \np129 (1977), 39; A. H. Chamseddine, \ap113
(1978), 212; S. MacDowell and F. Mansouri, \prl38
(1977), 739.}. In fact, Pagels' construction of general
relativity is trivially extendable to any dimension.

\newsection Two-dimensional gravity as gauge theory.

In two dimensions the curvature tensor $R_{\mu\nu\rho
\sigma}$ may be written in terms of the curvature
scalar $R$
$$
R_{\mu\nu\rho\sigma}={1\over 2}R(g_{\mu\rho}g_{\nu\sigma}
-g_{\mu\sigma}g_{\nu\rho})\numbereq\name{\eqinaut}
$$
so that $R$ alone completely characterizes the
local geometry. The relationship between the Ricci tensor
$R_{\mu\nu}$ and $R$ is such that the Einstein tensor
vanishes identically. The fact that Einstein gravity
is inappropriate in two dimensions could also be deduced
from the fact that the curvature term in the
Hilbert-Einstein action is the Euler characteristic
class for the manifold, which is independent of the
metric. Because the full information on space-time
geometry is contained
in the Ricci scalar $R$,
Jackiw \ref\jac{R. Jackiw {\it Quantum Theory of Gravity}
ed. S. Christensen {\it (Adam Hilger, Bristol)} (1984).}
and Teitelboim \ref\teit{C. Teitelboim {\it Quantum
Theory of Gravity} ed. S. Christensen {\it (Adam
Hilger, Bristol)} (1984).} proposed as an equation of
motion $R={\Lambda}$. This equation is derived from the
following action [\jac]
$$
S=\int d^2x{\sqrt g}N(R-{\Lambda}) \numbereq\name{\eqriv}
$$
where $N$ is an auxiliary scalar field. By varying this
action with respect to
$N$ and $g_{\mu\nu}$ we recover the equations of motion
$$
R={\Lambda}, \quad ({\nabla_{\mu}}{\nabla_{\nu}}-g_{\mu\nu}
{\nabla^2})N+{1\over 2}g_{\mu\nu}{\Lambda}N=0.
\numbereq\name{\eqmaur}
$$
Notice that the equation $R=\Lambda$ does not involve $N$,
this equation determines the metric,
while the other one determines $N$ with no
further restriction on $g_{\mu\nu}$. Despite appearances, this
model can be written as a $sl(2,R)$ gauge theory \ref\fuk{T.\tie
Fukuyama and K.\tie Kamimura, \pl160, (1985), 259;
M. Awada and A. H. Chamseddine, \pl233 (1989), 79;
A. H. Chamseddine and D. Wyler, \pl228 (1989), 75; \np340
(1990), 595; K. Isler and C. Trugenberger, \prl63 (1989), 834;
H. Terao, \np395 (1993), 623.}, in the same
way that Pagels wrote a gauge theoretic action for four dimensional
general relativity ( See also \ref{\jack}{ H. Verlinde in
{\it The Sixth Marcel Grossman Meeting on General
Relativity}, H. Sato ed. (World Scientific, 1992);
D. Cangemi and R. Jackiw,
\prl69 (1992), 233; \pl299 (1993), 24; D. Cangemi, R. Jackiw
and B. Zwiebach, \ap245 (1996), 408.} for a gauge
theoretical formulation of the dilaton gravity).
In fact Pagels' formulation may be extended to
any number of dimensions. We adopt the following conventions:
Greek indices  ${\mu}=0\ldots d-1$ refer to space-time; upper case
latin $A$ to an $SO(d,1)$ gauge group and lower case latin $a=0,\ldots,
d-1$ to the local Lorentz group, which is an $SO(d-1,1)$ subgroup
of the full
$SO(d,1)$ gauge group. We introduce $SO(d,1)$ gauge fields
${\omega_{\mu}^
{AB}}=-{\omega_{\mu}^{BA}}$ which transform as the adjoint
representation of $SO(d,1)$ and a scalar field in the fundamental
representation, ${\phi^A}$. The action, according to Pagels, is $$
S = \int d^dx\alpha \epsilon^{A\ldots DEF} D\phi^A\ldots R^{DE}\phi^F
+  \beta D\phi^A\ldots D\phi^E\phi^F.  \numbereq\name\eqpagels $$
Here, the elipsis refers to the appropriate number of $ D\phi$'s
to soak up the indices on the $SO(d,1)$ $ \epsilon$ symbol, $
D\phi^A$ is the covariant exterior derivative of the zero-form $
\phi^A$, and $ R^{AB}$ is the curvature two-form of the $SO(d,1)$
connection. Identifying the gravitational {\it vielbein\/} with
the part of $ D\phi^A$ transverse to $ \phi^A$, the spin-connection
with a similar transverse piece of the $SO(d,1)$ connection and
imposing the condition $ \phi^2 = M^2$ we recover the Hilbert-Einstein
action with non-zero cosmological constant. In two-dimensions this
procedure yields a total divergence.
Instead we shall use a variant of this action
$$
S={\int}d^2x{\epsilon^{\mu\nu}}{\epsilon^{ABC}}
R_{\mu\nu}^{AB}{\phi^C}\numbereq\name{\eqcompa}
$$
where the gauge field strength is given by
$R_{\mu\nu}^{AB}={\partial_{[\mu}}{\omega_{\nu]}^{AB}}+
{\omega_{[\mu}^{AC}}{\omega_{\nu]}^{CB}}$ and
$\epsilon^{ABC}$ is the antisymmetric $sl(2,R)$ symbol.

The three components of the $sl(2,R)$ gauge fields
$\omega_{\mu}^{AB}$ will eventually be identified as the
spin connection $\omega_{\mu}^{ab}$ and the two
components of the veirbein $\omega_{\mu}^{a3}=e_{\mu}^{a}$.
The equations of motion follow by varying
the action with respect to the dynamical
degrees of freedom $\omega_{\mu}^{AB}$ and ${\phi^A}$
$$
R^{AB}=0 \qquad D{\phi^C}=0\numbereq\name{\eqdivina}
$$
where $D{\phi^A}={\partial}{\phi^A}+{\omega^{AB}}{\phi^{B}}$
is the covariant $sl(2,R)$ derivative. The general solution
to the first equation can be written locally as $\omega^{AB}
=(g^{-1}dg)^{AB}=(g^{-1})^{AB}(dg)^{CB}$ where
$g^{AB}{\in sl(2,R)}$. Then the solution to the second
equation can be written as ${\phi^C}=(g^{-1})^{CD}{\phi_{0}^D}$,
where ${\phi_0}$ are constants. We observe that the space
of all solutions of the classical equations
of motion modulo gauge transformations is
finite dimensional. Thus the action (\eqcompa) describes
a topological field theory.

In order to make contact with two-dimensional gravity
we make the following identifications for the gauge
fields: ${\omega_{\mu}^{a3}}=s^{-1}e_{\mu}^a$,
where $e_{\mu}^a$ $a=1,2$ are the two components
of the veirbein and ${\omega_{\mu}^{ab}}={\omega_{\mu}}$
is the spin connection. The introduction
of the arbitrary scale parameter $s$ is necessary
because while the veirbein is
dimensionless, the gauge field has dimension one.
Then the equations $R_{\mu\nu}^{a3}=0$ provide
the vanishing of torsion and establish the
relation between the veirbein $e_{\mu}^a$ and
the spin connection ${\omega_{\mu}}$, $\omega_{\mu}=
-e^{-1}{\epsilon^{\mu\nu}}{\partial_{\nu}}e_{\rho}^a
e_{a\mu}$. The remaining equation gives us $R={\Lambda}$
with the identification $\Lambda=s^{-2}$. Finally
combining the two equations for $\phi^a$, we can
write the equation for $\phi^3$ as follows
$$
({\nabla_a}{\nabla_b}-g_{ab}{\nabla^2}){\phi^3}+
{1\over 2}s^{-2}g_{ab}{\phi^3}=0.\numbereq\name{\eqmalk}
$$
So by identifying $\phi^3$ with $N$ we have shown that
the $sl(2,R)$ gauge theory classically is equivalent
to the theory which was proposed by Jackiw [\jac] to
describe two-dimensional gravity.

\newsection $sl(2,R)$ Gauge.

The variation of $e_{\mu}^A$ under a $sl(2,R)$ gauge
transformation is
$$
{\delta}{e_{\mu}^A}=-D_{\mu}{\rho^A}=
-{\partial_{\mu}}{\rho^A}-{\epsilon^{ABC}}
e_{\mu}^{B}{\rho^C}\numbereq\name{\eqrutib}
$$
where $\rho^A$ is the parameter of transformation. Under a
diffeomorphism generated by a vector field $-\upsilon^{\kappa}$,
the standard transformation law would be
$$
{\delta}e_{\mu}^A=-{\upsilon^{\kappa}}{\partial_{\kappa}}
e_{\mu}^A-{\partial_{\mu}}{\upsilon^{\kappa}}e_{\kappa}^A.
\numbereq\name{\eqfghoi}
$$
The two-dimensional metric is $g_{\mu\nu}=e_{\mu}^{a}
e_{\nu}^{b}{\eta_{ab}}$, since we have identified
$e_{\mu}^a$ with the veirbein. We choose light-cone
coordinates and gauge fix in the following manner:
$e_{-}^{-}=0$, $e_{+}^{+}={h_{++}\over {\lambda}}$
$e_{+}^{-}={1\over {\lambda}}$,
$e_{-}^{+}=-{1\over 2}{\exp({\phi})}{\lambda}$
where $\phi(x)$, $h_{++}(x)$ and $\lambda(x)$
are arbitrary functions. With this particular
choice, the metric becomes
$$
g_{--}=0, \qquad g_{++}=h_{++}{\exp {\phi}},
\qquad g_{+-}=g_{-+}={1\over 2}{\exp {\phi}}.
\numbereq\name{\eqrw}
$$
This particular gauge was introduced in \ref\wen{K.-W. Xu and
C.-J. Zhu \imp6 (1991), 2331.} in order to
investigate the symmetry structure of two-dimensional
gravity.
Initially we have treated the veirbein $e_{\mu}^a$
and the spin connection ${\omega}_{\mu}^{ab}$ as
independent variables. The equations of motion (\eqdivina)
provide a relation between them. In the
gauge we have chosen,
with the particular choice for the components
of the veirbein, the two components of the spin-connection
read
$$
{\omega_{+}}=-{\partial_{-}}h_{++}-({\partial_{-}}{\phi})h_{++}
+{\lambda}{\partial_{+}}({1\over {\lambda}}), \qquad
{\omega}_{-}=-{\partial_{-}}{\phi}-{{\partial_{-}}
{\lambda}\over {\lambda}}.\numbereq\name{\eqwbvn}
$$
If we substitute these values into the equation
$R_{\mu\nu}^{ab}=0$, we find the following equation of motion
for the dynamical degrees of freedom $h_{++}(x)$ and $\phi(x)$
of the metric
$$
{\partial_{+}}{\partial_{-}}{\phi}+{{\Lambda}\over 2}
{\exp {\phi}}={\partial_{-}^2}h_{++}+({\partial_{-}^2}
{\phi})h_{++}+({\partial_{-}}{\phi})({\partial_{-}}h_{++}).
\numbereq\name{\eqbopiu}
$$
Under a gauge transformation
the veirbein (gauge field) $e^{-}_{-}$ transforms as
${\delta}e_{-}^{-}=-{\partial_{-}}{\rho^{-}}
+e_{-}^{-}{\rho^3}-{\omega_{-}}{\rho^{-}}$.
Then the residual $sl(2, R)$ gauge transformations correspond
to those ones which respect this particular gauge
choice, namely the transformations which satisfy ${\delta}e_{-}^{-}=0$.
The solution to this equation constrains $\rho^-$
in the form $\rho^{-}=-e^{-}_{+}{\tau}(x^+)$, with
$\tau(x^+)$ being an arbitrary function of $x^+$.
As a result all gauge transformations generated by
$\rho^+$, $\rho^3$ and $\rho^{-}=-e^{-}_{+}{\tau}(x^+)$
respect this particular gauge choice.
Let us proceed and  calculate the variation of
the physical degrees of freedom $\delta g_{++}$ and
$\delta g_{+-}$ under residual gauge transformations.
We find that
$$
\eqalign{
{\delta}g_{+-}=-{\delta}e^{-}_{+}e^{+}_{-}-e^{-}_{+}
{\delta}e^{+}_{-}&={1\over 2}e^{\phi}{\delta}{\phi}
={1\over 2}e^{\phi}{\partial_{+}}{\tau}(x^+)
+{1\over 2}e^{\phi}({\partial_{+}}{\phi})
{\tau}(x^+)\cr
&-{1\over 2}e^{\phi}({\partial_{-}}h_{++}){\tau}(x^+)
-{1\over 2}e^{\phi}({\partial_{-}}{\phi})h_{++}{\tau}(x^+)
-{\partial_{-}}
(e^{-}_{+}{\rho^{+}}). \cr}
\numbereq\name{\eqlygeros}
$$
Similarly we find that
$$
\eqalign{
&{\delta}g_{++}=-2{\delta}e^{+}_{+}e^{-}_{+}-2e^{+}_{+}
{\delta}e^{-}_{+}=e^{\phi}{\delta}{\phi}h_{++}+
e^{\phi}{\delta}h_{++}= \cr
&-2{\partial_{+}}(e^{-}_{+}
{\rho^{+}})-{\lambda}e^{\phi}({\partial_{+}}{\phi})
{\rho^{+}}+{\lambda}e^{\phi}({\partial_{-}}{\phi})
h_{++}{\rho^{+}}+{\lambda}e^{\phi}({\partial_{-}}h_{++})
{\rho^{+}} \cr
&+e^{\phi}h_{++}{\partial_{+}}{\tau}(x^+)+
e^{\phi}h_{++}{\partial_{+}}{\phi}{\tau}(x^+)-e^{\phi}
h_{++}{\partial_{-}}h_{++}{\tau}(x^+)-e^{\phi}h_{++}^2
{\partial_{-}}{\phi}{\tau}(x^+).\cr}
\numbereq\name{\eqkitsios}
$$
We can now further fix the gauge by selecting a specific form
for $e^{\phi}$. We choose $e^{\phi}={1\over ({\alpha}(x^+)
x^-+{\beta}(x^+))^2}$. We seek the gauge transformations
which preserve this form of $e^{\phi}$, namely we permit
the arbitrary functions ${\alpha}$ and ${\beta}$ to vary
in such a manner that preserve the form of $e^{\phi}$.
This will determine $\rho^+$ in terms of three unknown
functions ${\eta}(x^+)$, ${\delta}{\alpha}(x^+)$ and
${\delta}{\beta}(x^+)$ as follows
$$
{\rho^+}=-e^{+}_{-}{{\delta}{\alpha}\over {\beta}}(x^-)^2-
e^{+}_{-}{{\alpha}{\delta}{\beta}\over {\beta}^2}(x^-)^2-
e^{+}_{-}{{\delta}{\beta}\over {\beta}}x^{-}-e^{+}_{-}
{\eta}{\alpha^2}(x^-)^2-2e^{+}_{-}{\eta}{\alpha}{\beta}
x^{-}-e^{+}_{-}{\beta^2}{\eta}.\numbereq\name{\eqrougik}
$$
We have omitted the transformation which is generated
by $\tau(x^+)$ for simplicity. Having determined the form
of ${\rho^+}$ we will now determine the effect of the
transformations generated by ${\eta}(x^+)$,
${\delta}{\alpha}(x^+)$ and
${\delta}{\beta}(x^+)$ on the only remaining degree of
freedom $h_{++}$. The solution to (\eqbopiu) can be written
as
$$
h_{++}(x^{+}, x^{-})={\sqrt 2}J^{(+)}(x^{+})-2J^{(0)}
(x^{+})x^{-}+{\sqrt 2}J^{(-)}(x^+)(x^{-})^2\numbereq\name{\eqevans}
$$
where $J^{(-)}=
-2{\partial_{+}}{\alpha}+2{{\alpha}{\partial_{+}}{\beta}
\over {\beta}}+{{\Lambda}\over 2}$, $J^{(+)}$ and $J^{(0)}$ are
arbitrary functions of $x^{+}$. By substituting (\eqevans) and
(\eqrougik) into (\eqkitsios) we find that $J^{(-)}$, $J^{(+)}$
and $J^{(0)}$ transform as follows under the residual $sl(2, R)$
gauge transformations
$$
\eqalignno{{\delta}J^{(+)}&=-{1\over {\sqrt 2}}
{\partial_{+}}({\eta}{\beta}^2)+{\sqrt 2}{\eta}
{\beta}^2J^{(0)}+2({\eta}{\alpha}
{\beta}+{{\delta}{\beta}\over
{\beta}})J^{(+)}&{\global\advance\eqnumber by 1}
(\the\secnumber.\the\eqnumber a)\name{\eqvirj}\cr
{\delta}J^{(0)}&=-{\partial_{+}}
({\eta}{\alpha}{\beta}+{{\delta}{\beta}\over {\beta}})
+{\sqrt 2}{\eta}{\beta}^2J^{(-)}
-{\sqrt 2}({{\delta}{\alpha}\over {\beta}}+
{{\delta}{\beta}{\alpha}\over {\beta^2}}+{\eta}
{\alpha^2})J^{(+)}&
(\the\secnumber.\the\eqnumber b)\cr
{\delta}J^{(-)}&=-{1\over {\sqrt 2}}{\partial_{+}}
({\eta}{\alpha^2}+{{\delta}{\alpha}\over {\beta}}+{{\delta}
{\beta}{\alpha}\over {\beta^2}})-{\sqrt 2}({{\delta}{\alpha}\over
{\beta}}+{{\delta}{\beta}{\alpha}\over {\beta^2}}+{\eta}{\alpha^2})
J^{(0)}-2({\eta}{\alpha}{\beta}
+{{\delta}{\beta}\over {\beta}})J^{(-)}
(\the\secnumber.\the\eqnumber c)\cr}
$$
This is the main result of the paper.
The $J^{(+)}$, $J^{(-)}$ and $J^{(0)}$ transform under
the residual $sl(2,R)$ gauge transformations as the
adjoint of an $sl(2,R)$ Kac-Moody algebra. The parameters
of transformations $\epsilon^{+}={1\over {\sqrt 2}}
({\eta}{\beta}^2)$,
$\epsilon^{0}=({\eta}{\alpha}{\beta}+
{{\delta}{\beta}\over {\beta}})$ and
$\epsilon^{-}={1\over {\sqrt 2}}
({\eta}{\alpha^2}+{{\delta}{\alpha}\over {\beta}}+{{\delta}
{\beta}{\alpha}\over {\beta^2}})$
are part of the original $sl(2,R)$
gauge transformations with $\rho^+$ constrained
as in Eqn. (\eqrougik).
The Polyakov $sl(2,R)$ Kac-Moody algebra in light cone gauge
is a remnant of an $sl(2,R)$ Kac-Moody symmetry algebra which
manifests itself as a residual symmetry algebra in the
$sl(2,R)$ gauge and it is related to the original $sl(2,R)$
tangent space gauge symmetry. Upon gauge fixing the original
gauge transformations behave as coordinate transformations.
In the next section we will investigate the relation of the $sl(2, R)$
gauge to the light-cone and to the conformal gauge.

\newsection Relation to other Gauges.

Polyakov has proposed to study two-dimensional
gravity in the light-cone gauge where
the line element is
$$
ds^2=dx^{+}dx^{-}+g_{++}(x)(dx^{+})^2.\numbereq\name{\eqruto}
$$
The light-cone gauge is recovered by setting $\phi=0$
in (\eqrw).
Then the metric takes the following form:
$$
g_{++}=h_{++}(x), \qquad g_{--}=0, \qquad g_{+-}=g_{-+}
={1\over 2}\numbereq\name{\eqwdch}
$$
which is exactly the metric in the light cone gauge.
The condition that the residual gauge transformations
respect the gauge choice $\delta e_{-}^{-}=0$ and
$\delta g_{+-}=0$ provide the following constraints on
$\rho^+$ and $\rho^-$
$$
\rho^-=-e^{+}_{-}{\tau}(x^+), \qquad \rho^{+}=
-e^{+}_{-}h_{++}{\tau}(x^+)+e^{+}_{-}{\partial_{+}}
{\tau}(x^+)x^{-}-e^{+}_{-}{\eta}(x^+).\numbereq\name{\eqvoubn}
$$
We notice that $\rho^+$ is further constrained than before
Eqn. (\eqrougik).
This particular relation can also be recovered from (\eqrougik) by
setting $\alpha={\delta}{\alpha}={\delta}{\beta}=0$ and
$\beta=1$.

In light-cone gauge the equation of motion for $h_{++}$ becomes
${\partial_{-}^2}h_{++}(x)={{\Lambda}\over 2}$. This can be seen
by either setting $\phi=0$ into (\eqbopiu) or by
substituting the values for the components of the spin
connection in light-cone gauge
$$
{\omega_{+}}=-{\partial_{-}}h_{++}
+{\lambda}{\partial_{+}}({1\over {\lambda}}), \qquad
{\omega}_{-}=-{{\partial_{-}}
{\lambda}\over {\lambda}} \numbereq\name{\eqhgvn}
$$
into (\eqdivina).
The solution to the equation of motion for the
gravitational degree of freedom $h_{++}(x^+, x^-)$ can be
parametrised again in the following form
$$
g_{++}=h_{++}(x^{+}, x^{-})={\sqrt 2}J^{(+)}(x^{+})-2J^{(0)}
(x^{+})x^{-}+{\sqrt 2}J^{(-)}(x^+)(x^{-})^2 \numbereq\name{\eqvhdl}
$$
where $J^{(-)}={{\Lambda}\over 2}$. The residual
gauge transformations, Eqn. (\eqvoubn),
act on the gravitational
degree of freedom $h_{++}(x)$ as follows
$$
\eqalign{
{\delta}h_{++}(x)=-{\partial_{+}}
h&_{++}(x){\tau}(x^{+})
-{\partial_{+}}{\tau}(x^{+}){\left(2-x^{-}{\partial_{-}}
\right)}h_{++}(x)\cr
&+{\partial_{+}^2}{\tau}(x^{+})x^{-}
-{\partial_{+}}{\eta}(x^{+})
-{\eta}(x^{+})h_{++}(x). \cr}
\numbereq\name{\eqrjkp}
$$
This permits us to calculate the variations of the
currents $J^{(+)}(x^{+})$, $J^{(-)}(x^{+})$ and
$J^{(0)}(x^{+})$ under the residual
gauge transformations generated by $\eta(x^+)$
$$
{\delta}J^{(+)}=
-{1\over {\sqrt 2}}{\partial_{+}}{\eta}(x^{+})+
{\sqrt 2}J^{(0)}{\eta}(x^{+}),
\qquad {\delta}J^{(0)}=
{\sqrt 2}J^{(-)}{\eta}(x^{+}), \qquad
{\delta}J^{(-)}=0.
\numbereq\name{\eqbpolyi}
$$
As we have further fixed our $sl(2, R)$ gauge in order to recover
the light-cone gauge, only part of the $sl(2, R)$ Kac-Moody
remains as a residual symmetry, more specifically only
the transformations
generated by $\eta(x^+)$ respect the light cone gauge.
Let's now demonstrate that residual gauge transformations
act on $h_{++}(x)$ as residual diffeomorphisms. The latter are
generated by a vector field of the form
$$
\upsilon^{+}=\upsilon^{+}(x^+), \qquad \upsilon^{-}=
-({\partial_{+}}\upsilon^{+})x^{-}+g(x^+)\numbereq\name{\eqasz}
$$
This is the form of the diffeomorphisms that preserve the
light-cone gauge.
Under these transformations $h_{++}$ transforms as
$$
\eqalign{
{\delta}h_{++}(x)=-{\partial_{+}}
h&_{++}(x){\upsilon^{+}}(x^{+})
-{\partial_{+}}{\upsilon^{+}}(x^{+}){\left(2-x^{-}{\partial_{-}}
\right)}h_{++}(x)\cr
&+{\partial_{+}^2}{\upsilon^{+}}(x^{+})x^{-}
-{\partial_{+}}g(x^{+})
-g(x^{+})h_{++}(x) \cr}
\numbereq\name{\eqrdmp}
$$
which is the same as Eqn. (\eqrjkp).

We briefly comment on the relation of the $sl(2,R)$ gauge
and the conformal gauge. By setting $h_{++}=0$
in (\eqrw) we recover the
conformal gauge $g_{++}=g_{--}=0$ and $g_{+-}=g_{-+}=
{1\over 2}e^{\phi}$.
The equation of motion (\eqbopiu) becomes ${\partial_{+}}
{\partial_{-}}{\phi}
+{{\Lambda}\over 2}e^{\phi}=0$ which we recognise
as the Liouville equation. The residual gauge transformations
which respect this particular form follow by simply demanding
$\delta h_{++}=0$. This will suplement the relation $\rho^{-}=
e^{-}_{+}{\tau}(x^+)$ which follows from $\delta e^{-}_{-}=0$
with $\rho^{+}=e^{+}_{-}{\sigma}(x^-)$. The residual gauge
transformations generated by $\rho^+$ and $\rho^{-}$ act on
$g_{+-}$ as diffeomorphisms generated by $({\tau}(x^+), {\sigma}
(x^-))$ since under those $g_{+-}$ transforms as
${\delta}g_{+-}=-{\partial_{+}}{\tau}-{\partial_{-}}{\sigma}$.
This is identical to the action of residual diffeomorphisms in
the conformal gauge.

\newsection Conclusions.

Let us review the results presented in this paper.
Two-dimensional gravity exhibits a $sl(2, R)$ current algebra
in the light-cone gauge. It is the presence of this algebra
that makes the theory tractable although it's origin appears
mysterious. In this paper we offered an explanation about
the origin of this important algebra.
The point is that two-dimensional gravity can be formulated
as a $sl(2, R)$ gauge theory. Upon gauge fixing
the current algebra manifests itself as a residual
symmetry algebra. Further gauge fixing to light cone gauge
restricts the gauge freedom and only part of the
current algebra survives as a residual symmetry.
The residual gauge transformations act on the gravitational
degrees of freedom as residual diffeomorphisms \ref{\amp}
{A. M. Polyakov, \imp5 (1990), 833.}.

\newsection Acknowledgments.

I would like to thank A. Chamseddine and
M. Evans for useful discussions.
This work was supported in part by the Department of Energy Contract
Number DE-FG02-91ER40651-TASKB. 

\immediate\closeout1
\bigbreak\bigskip

\line{\twelvebf References. \hfil}
\nobreak\medskip\vskip\parskip

\input refs

\vfil\end